\documentclass[final,3p,times,twocolumn]{elsarticle}
\biboptions{comma,sort&compress}
\usepackage{ecrc}
\usepackage{here}
\usepackage{graphicx}
\usepackage{epsfig}
\usepackage{epstopdf}
\usepackage{amsmath}

\def\nin{\noindent}
\def\beq{\begin{equation}}
\def\eeq{\end{equation}}
\def\bea{\begin{eqnarray}}
\def\eea{\end{eqnarray}}
\def\nnb{\nonumber}
\def\la{\langle}
\def\ra{\rangle}
\def\ga{\left(}
\def\dr{\right)}

\usepackage{graphicx}
\usepackage{here}
\def\beq{\begin{equation}}
\def\eeq{\end{equation}}
\def\bea{\begin{eqnarray}}
\def\eea{\end{eqnarray}}
\def\bq{\begin{quote}}
\def\eq{\end{quote}}
\def\ve{\vert}
\def\nnb{\nonumber}
\def\ga{\left(}
\def\dr{\right)}

\def\nnb{\nonumber}
\def\la{\langle}
\def\ra{\rangle}
\def\nin{\noindent}
\def\ba{\begin{array}}
\def\ea{\end{array}}

\def\als{\alpha_s}

\def\gg2{ \la\alpha_s G^2 \ra}
\def\gg3{g^3f_{abc}\la G^aG^bG^c \ra}
\def\ggg4{\la\als^2G^4\ra}

\def\beq{\begin{equation}}
\def\enq{\end{equation}}
\def\beqa{\begin{eqnarray}}
\def\enqa{\end{eqnarray}}
\def\nnb{\nonumber}

\def\MeV{\nobreak\,\mbox{MeV}}
\def\GeV{\nobreak\,\mbox{GeV}}
\def\keV{\nobreak\,\mbox{keV}}

\newcommand{\rag}{\rangle}
\newcommand{\lag}{\langle}

\def\ln{\mbox{Log}}
\def\gg{\lag g^{2}_{s} G^2 \rag}
\def\ggg{\lag g^{3}_{s}G^3\rag}

\volume{00}
\firstpage{1}
\journalname{Nuclear and Particle Physics Proceedings }
\runauth{D. Rabetiarivony}
\jnltitlelogo{Nuclear and Particle Physics Proceedings }

\begin{document}
\begin{frontmatter}

\title{Doubly hidden $0^{++}$ molecules and tetraquarks states from QCD at NLO \tnoteref{text1}}

\author[label1]{R. M. Albuquerque}
\ead{raphael.albuquerque@uerj.br}
\address[label1]{Faculty of Technology, Rio de Janeiro State University (FAT,UERJ), Brazil.}

\author[label2,label3]{S. Narison}
\ead{snarison@yahoo.fr}
\address[label2]{Laboratoire Univers et Particules de Montpellier, CNRS-IN2P3, Case 070, Place Eug\`ene Bataillon, 34095 - Montpellier, France.}
\address[label3]{Institute of High Energy Physics of Madagascar (iHEPMAD), University of Ankatso, Antananarivo 101, Madagascar.}

\author[label3]{D. Rabetiarivony\fnref{fn1}}
\fntext[fn1]{Speaker}
\ead{rd.bidds@gmail.com} 

\author[label3]{G.~Randriamanatrika}
\ead{artesgaetan@gmail.com}

\tnotetext[text1]{Talk given at QCD20 International Conference (27--30 October 2020, Montpellier--FR)}

\pagestyle{myheadings}
\markright{ }

\begin{abstract}
\noindent
Motivated by the LHCb-group discovery of exotic hadrons in the range (6.2 $\sim$ 6.9) GeV, we present new results for the masses and couplings of $0^{++}$ fully heavy $(\bar{Q}Q)(Q\bar{Q})$ molecules and $(QQ)(\overline{QQ})$ tetraquaks  states from relativistic QCD Laplace Sum Rule (LSR) within stability criteria where Next-to-Leading Order (NLO) Factorized (F) Perturbative (PT) corrections is included. As the Operator Product Expansion (OPE) usually converges for $d\leqslant 6-8$, we evaluated the QCD spectral functions at Lowest Order (LO) of PT QCD and up to $\lag G^3 \rag$.
We also emphasize the importance of PT radiative corrections for heavy quark sum rules in order to justify the use of the running heavy quark mass value in the analysis. We compare our predictions in Table \ref{tab:results} with the ones from ratio of Moments (MOM). The broad structure arround (6.2 $\sim$ 6.9) GeV can be described by the $\overline{\eta}_c\eta_c$, $\overline{J/\psi}J/\psi$ and $\overline{\chi}_{c1}\chi_{c1}$ molecules or/and $\overline{S}_c S_c$, $\overline{A}_c A_c$ and $\overline{V}_c V_c$ tetraquarks lowest mass ground states. The narrow structure at (6.8 $\sim$ 6.9) GeV if it is a $0^{++}$ state can be a $\overline{\chi}_{c0}\chi_{c0}$ molecules or/and its analogue $\overline{P}_c P_c$ tetraquark. The $\overline{\chi}_{c1}\chi_{c1}$ predicted mass is found to be below the $\chi_{c1}\chi_{c1}$ threshold while for the beauty states, all of the estimated masses are above the $\eta_b \eta_b$ and $\Upsilon(1S)\Upsilon(1S)$ threshold.

\end{abstract} 
\scriptsize
\end{frontmatter}
\section{Introduction}
\begin{figure}[hbt] 
\begin{center}
{\includegraphics[scale=0.7]{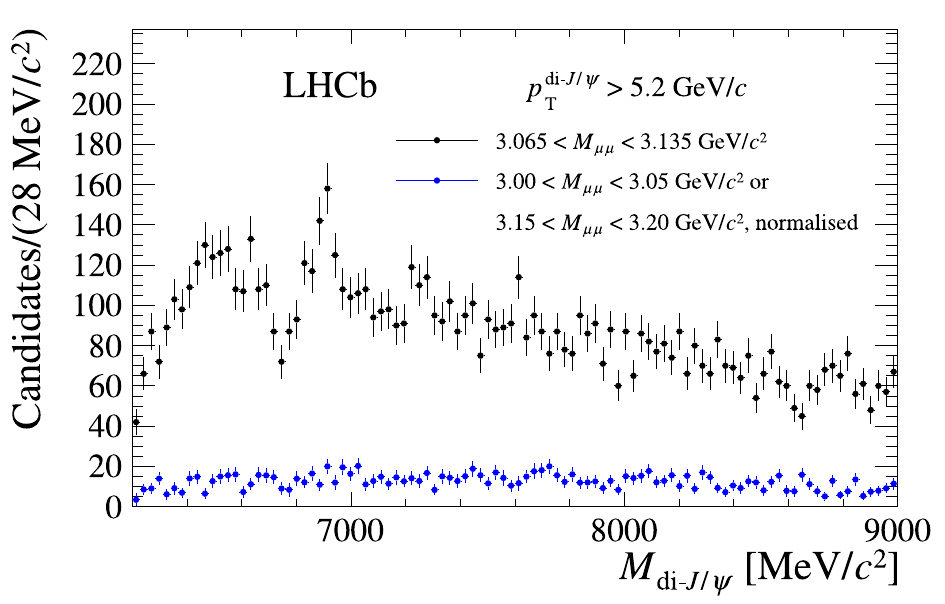}}
\caption{
{\scriptsize 
Ivariant mass spectrum of $J/\psi$-pair candidate. Data (resp. background regions) in black (resp. blue) signal\cite{LHCB1}.}
}
\label{fig:lhcb} 
\end{center}
\end{figure}

Recently, the LHCb collaboration \cite{LHCB1,LHCB2} studied the $J/\psi$-pair invariant mass spectrum and observed a narrow structure at $6.9$ GeV and a bump around $(6.2 \sim 6.7)$ GeV as we can see in Fig. \ref{fig:lhcb}. In Ref.\,\cite{QQQQ}, which we partly review here, we use the inverse Laplace Transform (LSR) \cite{BELL,BNR,BERT,NEUF,SNR} of QCD spectral sum rules (QSSR)\footnote{For reviews, see \cite{SVZa,Za,SNB1,SNB2,SNB3,CK,YND,PAS,RRY,IOFF,DOSCH}} to estimate the masses and couplings of $0^{++}$ fully heavy molecules and tetraquarks states for interpreting these recent experimental data. In so doing, we include the NLO PT corrections from factorized part diagrams which is a good approximation as we shall see that the contribution from Non-Factorized (NF) diagrams is almost negligible compared to the total $\alpha_s$ contributions. This feature has been already observed explicitly in our previous works \cite{AFNR,SNX2,ANRR,SU3}. We evaluate the four-quark correlators at LO of PT QCD up to the triple gluon condensate. 

\section{The Laplace sum rule}
We shall work with the finite energy version of the QCD inverse Laplace sum rules and their ratios:
\bea
\mathcal{L}^{c}_{n}(\tau ,\mu)&=&\int^{t_c}_{16 m^{2}_{Q}}dt\, t^n e^{-t\tau}\frac{1}{\pi}\mbox{Im}\Pi^{H}_{\mathcal{M},\mathcal{T}}(t,\mu),\nnb\\
\mathcal{R}^{c}_{n}(\tau)&=&\frac{\mathcal{L}^{c}_{n+1}}{\mathcal{L}^{c}_{n}},
\label{eq:LSR}
\eea
where $m_Q$ is the heavy quark mass, $\tau$ is the LSR variable, $n=0,1$ is the degree of moments, $t_c$ is the "QCD continuum" which parametrizes, from the discontinuity of the Feynman diagrams, the spectral function $\mbox{Im}\Pi^{S}_{\mathcal{M},\mathcal{T}}(t,m^{2}_{Q},\mu^2)$ where $\Pi^{S}_{\mathcal{M},\mathcal{T}}(t,m^{2}_{Q},\mu^2)$ is the scalar correlator defined as:
\beq
\Pi^{S}_{\mathcal{M},\mathcal{T}}(q^2) = \hspace*{-0.2cm} \int \hspace*{-0.1cm} d^4 x\, e^{-i q x}\lag 0 \ve \mathcal{T} \mathcal{O}^{S}_{\mathcal{M},\mathcal{T}}(x)(\mathcal{O}^{S}_{\mathcal{M},\mathcal{T}}(x))^{\dag} \ve 0 \rag,
\eeq
where $\mathcal{O}^{S}_{\mathcal{M},\mathcal{T}}(x)$ are the interpolating currents for the molecule $\mathcal{M}$ and tetraquark $\mathcal{T}$ states. The superscript $S$ refers to the spin of the scalar particles.
\vspace*{-0.3cm}
\section{Interpolating currents}
\nin
We shall be concerned with the interpolating currents given in Eq. \ref{eq:current} and Table \ref{tab:current}.
\bea
\mathcal{O}^{S}_{\mathcal{T}}=\epsilon_{abc}\,\epsilon_{dec}\,( Q^{T}_{a}\, C\,\gamma_\mu\, Q_b ) (\bar{Q}_d\, \gamma^\mu\,C\, \bar{Q}^{T}_{e})
\label{eq:current}
\eea
\vspace*{-0.8cm}
\begin{table}[hbt]
\setlength{\tabcolsep}{0.8pc}
\caption{{\scriptsize Interpolating currents $\mathcal{O}^{S}_{\mathcal{M},\mathcal{T}}$ with a definite C-parity describing the molecules and tetraquarks states. $Q,q\equiv c,b $}
}
\begin{center}
{\scriptsize
\begin{tabular}{ll}
\hline
\hline
Scalar States ($0^{++}$)&Current ($\mathcal{O}^{S}_{\mathcal{M},\mathcal{T}}$) \\
\hline
Molecules&\\
$\overline{\chi}_{q0} \, \chi_{q0}$ &
$( \bar{Q}\, Q ) (\bar{Q}\, Q)$ \\
$\overline{\eta}_q \, \eta_q$ &
$( \bar{Q}\, \gamma_5\, Q ) (\bar{Q}\, \gamma_5\, Q)$ \\
$\overline{J/\psi} \, J/\psi$, $\overline{\Upsilon} \, \Upsilon$ &
$( \bar{Q}\, \gamma_\mu\, Q ) (\bar{Q}\, \gamma^\mu\, Q)$ \\
$\overline{\chi}_{q1} \, \chi_{q1}$&
$( \bar{Q}\, \gamma_\mu \gamma_5\, Q ) (\bar{Q}\, \gamma^\mu \gamma_5\, Q)$ \\
Tetraquarks&\\
$\overline{P}_q \, P_q$ &
$( Q^{T}_{a}\, C\, Q_b ) (\bar{Q}_a\, C\, \bar{Q}^{T}_{b})$ \\
$\overline{S}_{q} \, S_{q}$ &
$( Q^{T}_{a}\,C\,\gamma_5 Q_b ) (\bar{Q}_a\,\gamma_5\,C\,\bar{Q}^{T}_{b})$ \\
$\overline{A}_q \, A_q$ &
$( Q^{T}_{a}\, C\,\gamma_\mu\, Q_b ) (\bar{Q}_a\, \gamma^\mu\,C\, \bar{Q}^{T}_{b})$ \\
$\overline{V}_q \, V_{q}$&
$( Q^{T}_{a}\, C\,\gamma_\mu \gamma_5\, Q_b ) (\bar{Q}_a\, \gamma^\mu \gamma_5\,C\, \bar{Q}^{T}_{b})$ \\
\hline\hline
\end{tabular}
}
\end{center}
\label{tab:current}
\end{table}

\vspace*{-0.5cm}
\section{The Spectral function}
\nin
We shall use the Minimal Duality Ansatz (MDA) for parametrizing the molecule spectral function:
{\small
\beq
\frac{1}{\pi}\mbox{Im}\Pi^{S}_{\cal{M}}\hspace*{-0.1cm}\simeq \hspace*{-0.1cm} f^2_{\cal{M}}M^8_{\cal{M}}\delta(t-M_{\cal{M}}^2)+\mbox{"QCD continuum"}\theta (t-t_c),
\label{eq:mda}
\eeq
}
where the "QCD continuum" is the imaginary part of the QCD correlator from the threshold $t_c$. The decay constant $f_{\cal{M}}$ (analogue to $f_{\pi}$) for the molecule state is defined as:
\bea
\lag 0 \ve \mathcal{O}^{S}_{\mathcal{M}} \ve \mathcal{M} \rag = f^{S}_{\mathcal{M}}\, M^{4}_{\mathcal{M}}.
\label{eq:coupling}
\eea
Interpolating currents constructed from bilinear (pseudo)scalar currents are not renormalization group invariants such that the corresponding decay constants possess anomalous dimension:
\bea
f^{S}_{\mathcal{M}}(\mu)=\hat{f}^{S}_{\mathcal{M}}(-\beta_1 \,a_s)^{4/\beta_1}(1-k_f\,a_s),
\eea
where $\hat{f}^{S}_{\mathcal{M}}$ is the renormalization group invariant coupling and $[-\beta_1=(1/2)(11-2n_f/3)]$ is the first coefficient of the QCD $\beta$-function for $n_f$ flavors. $a_s\equiv (\alpha_s/\pi)$ is the QCD coupling. $k_f=2.028(2.352)$ for $n_f=4(5)$ flavors.\\
\nin
Within such a parametrization, one obtains:
\bea
\mathcal{R}^{c}_{0}\equiv \mathcal{R}\simeq M^{2}_{\mathcal{M}},
\eea
where $M_{\mathcal{M}}$ is the lowest ground state mass. Analogous definitions can be obtained for the tetraquark states by changing the subscripts $\mathcal{M}$ into $\mathcal{T}$.
\section{NLO PT corrections and stability criteria}
\label{sec:conv}
\nin
Assuming a factorization of the four-quark interpolating current,
we can write the corresponding spectral function as a convolution of the two ones associated to two quark bilinear currents.
In this way, we obtain \cite{PICH,NPIV}:

\bea
\frac{1}{ \pi}{\rm Im} \Pi^{H}_{\mathcal{M},\,\mathcal{T}}(t)\hspace*{-0.3cm}&=&\hspace*{-0.3cm}\theta (t-16 M_Q^2)  \ga \frac{k}{ 4\pi} \dr ^2  t^2 \int_{4 m_Q^2}^{(\sqrt{t}-2m_Q)^2} dt_1 \nnb\\
&&\times \int_{4m_Q^2}^{(\sqrt{t}-\sqrt{t_1})^2} dt_2 \, \lambda^{1/2}\, \mathcal{K}^{H},
\label{eq:conv}
\eea
where $k$ is an appropriate normalization factor, $m_Q$  the on-shell heavy quark mass and
{\small
\bea
\mathcal{K}^{S,P}\hspace*{-0.3cm}&\equiv &\hspace*{-0.3cm}\ga  \frac{t_1}{ t}\hspace{-0.1cm}+\frac{t_2}{ t}-1  \dr ^2\times \frac{1}{ \pi}{\rm Im}\psi^{S,P}(t_1) \frac{1}{ \pi} {\rm Im} \psi^{S,P}(t_2)\,,\\
\mathcal{K}^{V,A}\hspace*{-0.3cm}&\equiv &\hspace*{-0.3cm}\left[\ga\frac{t_1}{ t}\hspace*{-0.1cm}+\frac{t_2}{ t}\hspace*{-0.1cm}-1\dr ^2\hspace*{-0.1cm}+\hspace*{-0.1cm}\frac{8 t_1 t_2}{t^2}\right] \times \frac{1}{ \pi}{\rm Im}\Pi^{V,A}(t_1) \frac{1}{ \pi} {\rm Im} \Pi^{V,A}(t_2)\,,\nnb
\eea
}
with the phase space factor:
{\small
\beq
\lambda=\ga 1-\frac{\ga \sqrt{t_1}- \sqrt{t_2}\dr^2}{ t}\dr \ga 1-\frac{\ga \sqrt{t_1}+ \sqrt{t_2}\dr^2}{ t}\dr~.
\eeq
}
The NLO expressions of the spectral functions of the bilinear equal masses (pseudo)scalar and (axial-)vector are known in the literature \cite{SNB1,SNB2,RRY,DJB}.\\
\nin The variables $\tau,\mu$ and $t_c$ are, in principle, free external parameters. We shall use stability criteria with respect to these free 3 parameters to extract the lowest ground state mass and coupling (more detailed discussions can be seen in \cite{AFNR,SNX2,ANRR,SU3,SNX1,ANR1,ANR2,ADKT} and references therein).
\vspace*{-0.3cm}
\section{The On-shell and $\overline{\mbox{MS}}$-scheme}
\nin In our analysis, we replace the on-shell (pole) masses $m_Q$ appearing in the spectral functions with the running masses $\overline{m}_Q(\mu)$ using the relation, to order $\alpha^2_s$\,\cite{TARR,COQ,BIN,SNB4}:
{\small
\bea
m_Q \hspace{-0.3cm}&=&\hspace{-0.3cm} \overline{m}_Q(\mu)\Big{[}1+\frac{4}{3} a_s+ (16.2163 -1.0414 n_l)a_s^2\nnb\\
&&\hspace{-0.3cm}+\ln{\ga\frac{\mu}{ m_Q}\dr^2} \ga a_s+(8.8472 -0.3611 n_l) a_s^2\dr\nnb\\
&&\hspace{-0.3cm}+\ln^2{\ga\frac{\mu}{ m_Q}\dr^2} \ga 1.7917 -0.0833 n_l\dr a_s^2...\Big{]},
\label{eq:msb}
\eea
}
for $n_l$ light flavours where $\mu$ is the arbitrary subtraction scale.
\vspace*{-0.3cm}
\section{QCD input parameters}
\nin
The QCD parameters which shall appear in the following analysis will be the QCD coupling $\alpha_s$, the charm and bottom quark masses $m_{c,b}$, the gluon condensates $ \lag\alpha_sG^2\rag \equiv \la \alpha_s G^a_{\mu\nu}G_a^{\mu\nu}\ra$ 
and $ \la g^3G^3\ra \equiv \la g^3f_{abc}G^a_{\mu\nu}G^b_{\nu\rho}G^c_{\rho\mu}\ra$. Their values are given in Table \ref{tab:param}.
\begin{table}[hbt]
\setlength{\tabcolsep}{.25pc}
 \caption{{\scriptsize QCD input parameters from recent QSSR analysis based on stability criteria.}} 
{\footnotesize
 {\begin{tabular}{@{}llll@{}}
&\\
\hline
\hline
Parameters&Values&Sources& Ref.    \\
\hline
$\alpha_s(M_Z)$ & $0.1181(16)(3)$ & $M_{\chi_{0c,b}}-M_{\eta_{c,b}}$ & LSR \cite{SNA}\\
$\overline{m}_c(\overline{m}_c)$ & $1286(16)$ MeV & $B_c \oplus J/\psi$ & Mom \cite{SNB6}\\
$\overline{m}_b(\overline{m}_b)$ & $4202(8)$ MeV & $B_c \oplus \Upsilon$ & Mom \cite{SNB6}\\
$\la\alpha_s G^2\ra\times 10^2$ & $(6.35\pm 0.35)$ GeV$^2$ & Hadrons & Average \cite{SNA}\\
$\la g^3  G^3\ra / \la\alpha_s G^2\ra$ & $(8.2\pm 2.0)$ GeV$^2$ & $J/\psi$ family & QSSR \cite{SNB8}\\
\hline\hline
\end{tabular}}
}
\label{tab:param}
\end{table}
\vspace*{-0.55cm}
\section{Molecules and tetraquarks states}
\nin
We shall study the charm channels and their beauty analogue. As the analysis will be performed using the same techniques, we shall illustrate it in the case of $\chi_{c0}\chi_{c0}$. The results are compiled in Tables \ref{tab:results}.
\subsection{$f_{\chi_{c0}\chi_{c0}}$ and $M_{\chi_{c0}\chi_{c0}}$}
\begin{figure}[hbt] 
\begin{center}
{\includegraphics[width=3.8cm,height=2.4cm]{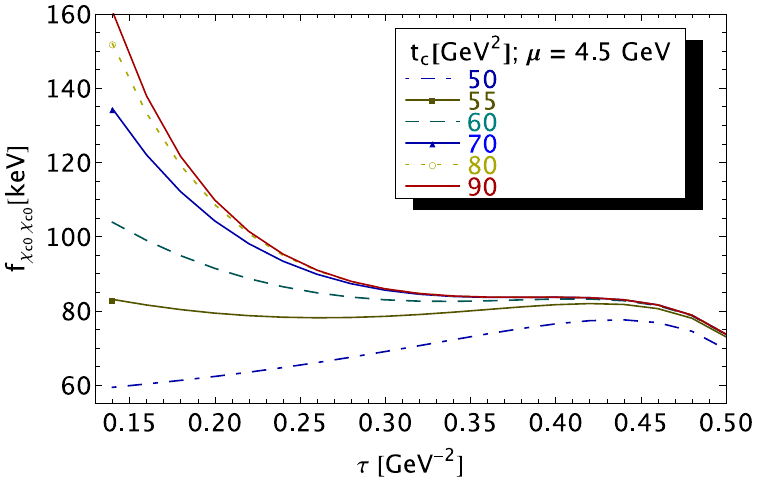}}
{\includegraphics[width=3.8cm,height=2.4cm]{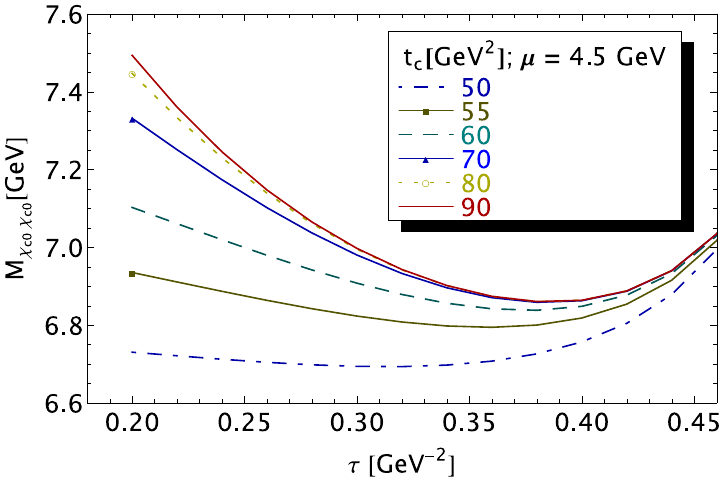}}
\caption{
{\scriptsize 
The coupling $f_{\chi_{c0}\chi_{c0}}$ and mass $M_{\chi_{c0}\chi_{c0}}$ at NLO as function of $\tau$ for different values of $t_c$, for $\mu=4.5$ GeV  and for the QCD parameters in Table\,\ref{tab:param}.}
}
\label{fig:chic0-nlo} 
\end{center}
\end{figure}

\nin
We study the behavior of the coupling and mass in term of the LSR variable $\tau$ for different values of $t_c$ at NLO as shown in Fig.\,\ref{fig:chic0-nlo}.
\nin
 We consider as final results the mean of the value corresponding to the beginning of $\tau$ stability for $[t_c\,(\GeV),\,\tau\,(\GeV^{-2})]\simeq [55,\,0.36]$  and the one where the $t_c$ stability is reached for $[t_c\,(\GeV),\,\tau\,(\GeV^{-2})]\simeq [70,\,0.38]$.
\subsection{$\mu$-stability}
\begin{figure}[hbt] 
\begin{center}
{\includegraphics[width=3.8cm,height=2.4cm]{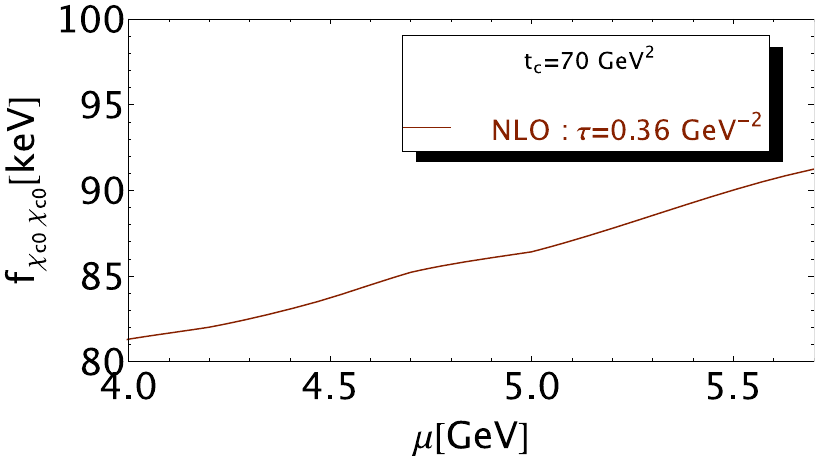}}
{\includegraphics[width=3.8cm,height=2.4cm]{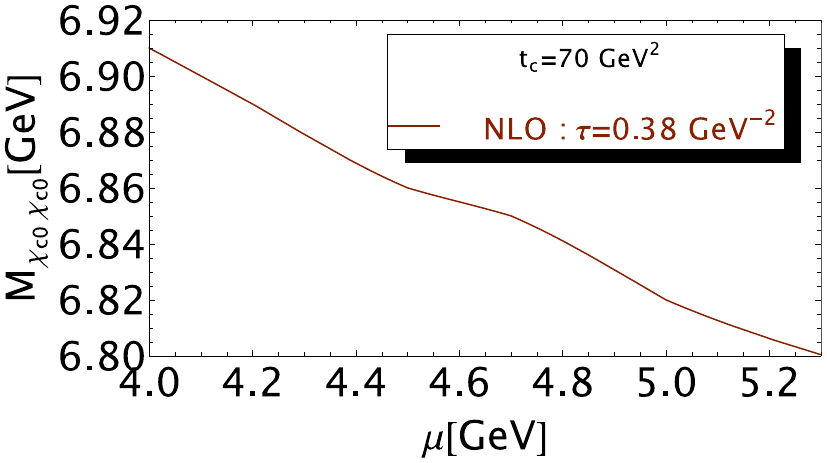}}
\caption{
{\scriptsize 
$\mu$-behavior of $f_{\chi_{c0}\chi_{c0}}$ and $M_{\chi_{c0}\chi_{c0}}$ for $t_c=70$ GeV$^2$ at NLO.}
}
\label{fig:chic0-mu} 
\end{center}
\end{figure}

\nin
Using the fact that the final results must be independent of the arbitrary parameter $\mu$, we consider as optimal result the one at the inflexion point for $\mu\simeq 4.5$ GeV (Fig.\,{\ref{fig:chic0-mu}}).
\vspace*{-0.5cm}
\subsection{The Factorization assumption}
\label{sec:fnf}
\begin{figure}[hbt] 
\begin{center}
{\includegraphics[width=3.7cm,height=2.6cm]{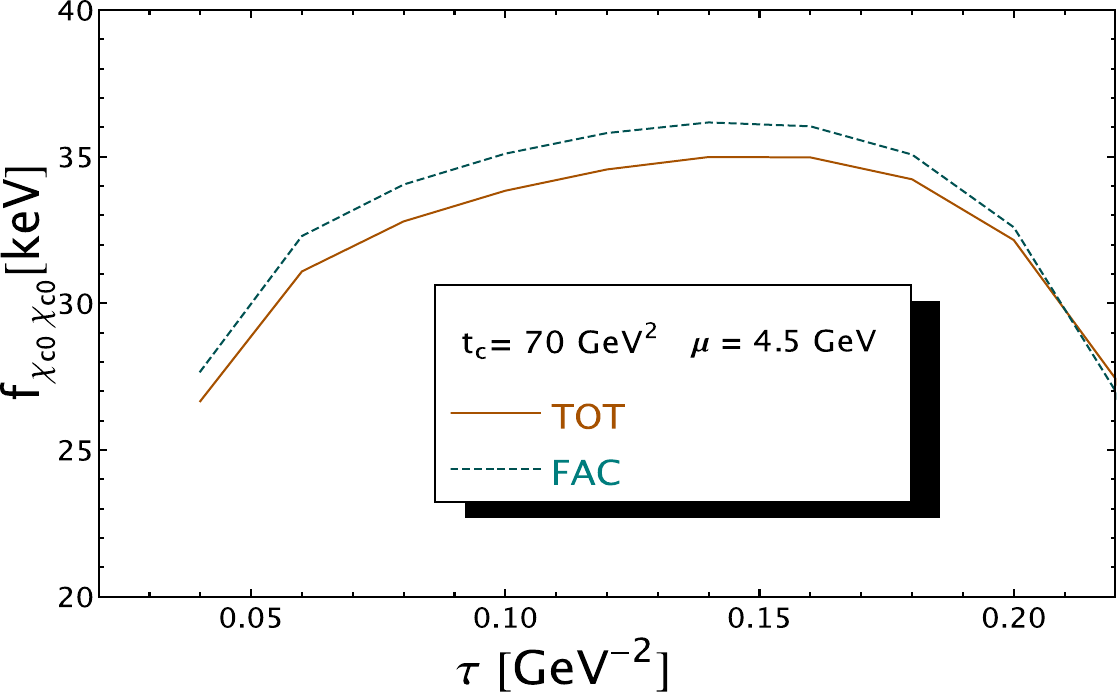}}
{\includegraphics[width=3.95cm,height=2.6cm]{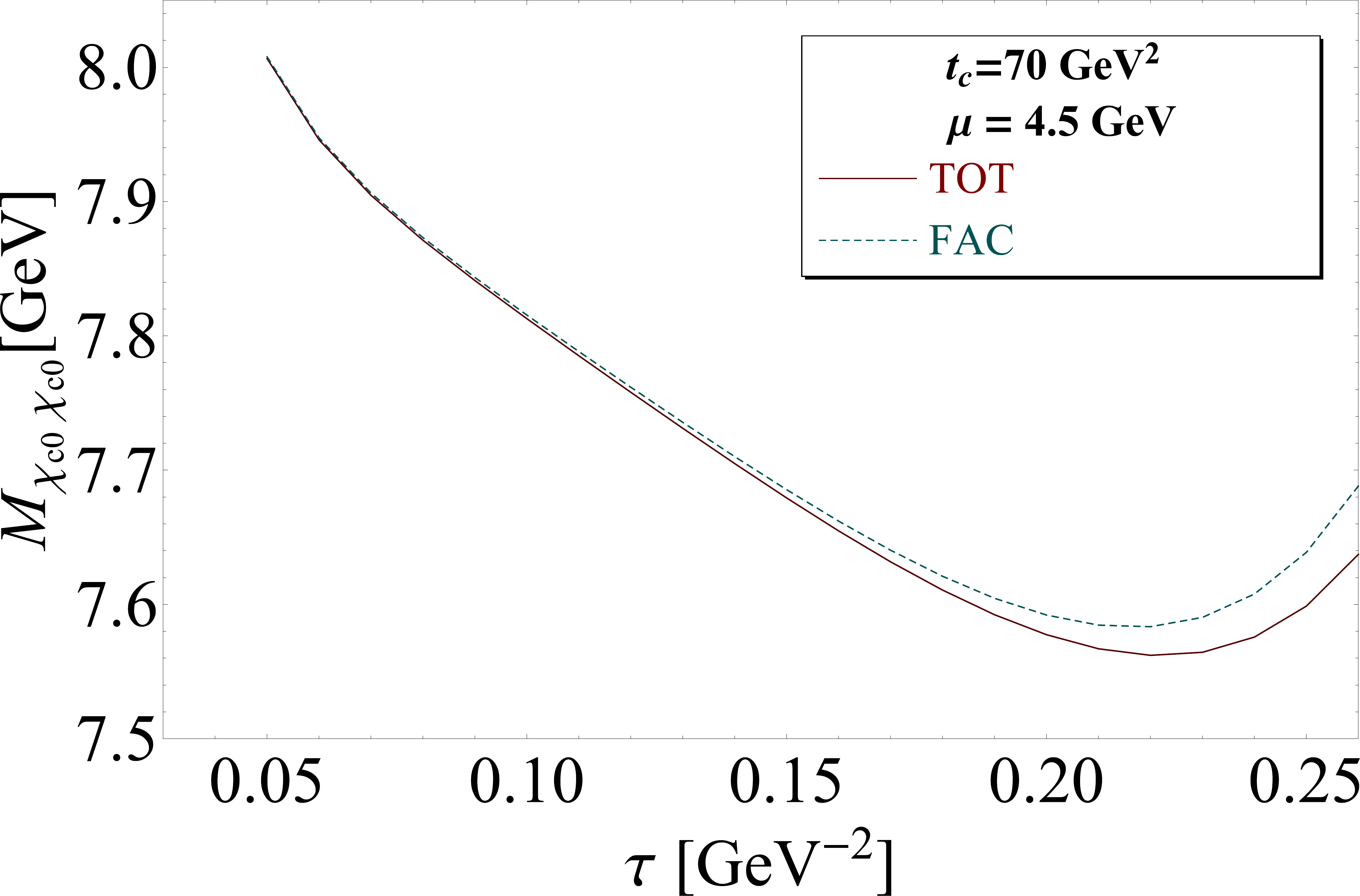}}
\caption{
{\scriptsize 
Comparison of the factorized (FAC) and F$\oplus$NF (TOT) at LO including $\lag \alpha_s G^2 \rag$ contribution for $t_c=70$ GeV$^2$ and $\mu=4.5$ GeV.}
}
\label{fig:chic0-fnf} 
\end{center}
\end{figure}

We have shown explicitly in \cite{QQQQ} that the contributions from the non-factorized diagrams appear at LO of perturbative series and for the $\lag \alpha_s G^2 \rag$ contributions. However, as we can see in Fig. \ref{fig:chic0-fnf}, the effect of these non-factorized diagrams is relatively small (about $1/(10N_c)$) compared to the total $F\oplus NF$ contributions. This feature justifies our approximation by using only the factorized part diagrams in the NLO perturbative contributions (see Section \ref{sec:conv}).
\subsection{The PT series} 
\nin
At LO, the two definitions of the quark mass lead to different predictions while at NLO this ambiguity between the running and pole quark mass definition is avoided. From the predictions for the running mass \cite{QQQQ} the effect of the PT corrections can be parametrized numerically as:
\bea
f_{\chi_{c0}\chi_{c0}}&\approx & 43~\mbox{keV}\,(1+8.7a_s\pm 7.57a^{2}_{s})~,\nnb\\
M_{\chi_{c0}\chi_{c0}}&\approx & 7.76~\mbox{GeV}\,(1-0.5a_s\pm 0.25a^{2}_{s})~,
\label{eq:ptser}
\eea
where the $a^{2}_{s}$ contributions have been estimated from a geometric growth of the PT coefficients \cite{NZ} and considered as an estimate of the uncalculated higher order terms of the PT series. One can notice from Eq.\,\ref{eq:ptser} that the PT series converge numerically but induce a relatively large systematic error for the coupling.
\nin
\begin{table*}[hbt]
\setlength{\tabcolsep}{0.385pc}
 \caption{{\scriptsize $0^{++}$ fully heavy molecules/tetraquarks couplings and masses predictions from LSR at NLO. The errors from QCD input parameters are from Table \ref{tab:param}. $\ve \Delta \mu\ve=0.20$ (resp. 0.25) GeV for the charm (resp. beauty) case. $\ve \Delta \tau\ve =0.02$ GeV$^{-2}$. In the case of asymetric errors, we take the mean value.}}
{\scriptsize{
\begin{tabular*}{\textwidth}{@{}ll ll  ll  ll ll ll ll ll ll l@{\extracolsep{\fill}}l}
\hline
\hline
                \bf Observables &\multicolumn{2}{c}{$\Delta t_c$}
					&\multicolumn{2}{c}{$\Delta \tau$}
					&\multicolumn{2}{c}{$\Delta \mu$}
					&\multicolumn{2}{c}{$\Delta m$}
					&\multicolumn{2}{c}{$\Delta \alpha_s$}
					&\multicolumn{2}{c}{$\Delta \alpha_s G^2$}
					&\multicolumn{2}{c}{$\Delta G^3-OPE$}
					&\multicolumn{2}{c}{$\Delta HO-PT$}
					&\multicolumn{2}{c}{Values}\\
\cline{2-3} \cline{4-5}\cline{6-7}\cline{8-9}\cline{10-11}\cline{12-13}\cline{14-15}\cline{16-17}\cline{18-19}
$q\equiv c,b$
                 & \multicolumn{1}{c}{c}
                 & \multicolumn{1}{c}{b} 
                 & \multicolumn{1}{c}{c} 
                 & \multicolumn{1}{c}{b} 
                 & \multicolumn{1}{c}{c} 
                 & \multicolumn{1}{c}{b}
		         & \multicolumn{1}{c}{c} 
                 & \multicolumn{1}{c}{b}
                 & \multicolumn{1}{c}{c} 
                 & \multicolumn{1}{c}{b}
                 & \multicolumn{1}{c}{c} 
                 & \multicolumn{1}{c}{b}
                 & \multicolumn{1}{c}{c} 
                 & \multicolumn{1}{c}{b}
                 & \multicolumn{1}{c}{c} 
                 & \multicolumn{1}{c}{b}
                 & \multicolumn{1}{c}{c} 
                 & \multicolumn{1}{c}{b}
                  \\
\hline
$f_{_{\mathcal{M},\mathcal{T}}}$[$\keV$]&&&&&&&&&&&&&&&&&&\\
$0^{++}$ Molecule &&&&&&&&&&&&&&&&&&\\
$\overline{\chi}_{q0}\chi_{q0}$&2.80&0.01&0.40&0.10&2.50&0.10&3.70&0.50&3.50&0.70&1.20&0.10&11.5&0.60&16.0&0.20&69$\pm$21&4.0$\pm$1.1\\
$\overline{\eta}_q \eta_q$&0.80&0.40&0.20&0.10&3.0&0.20&10.0&1.20&5.0&2.0&0.70&0.10&12.2&0.80&0.90&0.20&56$\pm$17&9.8$\pm$2.4\\
$\overline{J\psi}J\psi,\,\overline{\Upsilon}\Upsilon$&4.60&0.60&1.0&0.60&2.0&0.10&10.7&4.30&19.0&2.50&3.40&0.40&45.6&3.80&0.40&0.0&160$\pm$51&23.4$\pm$6.3\\
$\overline{\chi}_{q1}\chi_{q1}$&0.90&1.60&1.10&0.90&0.90&0.20&6.0&3.0&9.0&4.80&10.0&0.0&4.0&3.0&5.0&19.0&162$\pm$16&48.9$\pm$20.1\\
$0^{++}$ Tetraquark &&&&&&&&&&&&&&&&&&\\
Table \ref{tab:current} &&&&&&&&&&&&&&&&&&\\
$\overline{P}_q P_q$&1.40&1.80&0.40&2.30&3.40&0.50&7.20&1.0&3.50&1.0&1.30&0.10&8.90&3.50&4.80&1.20&60$\pm$14&6.5$\pm$4.9\\
$\overline{S}_q S_q$&0.10&0.10&0.70&0.20&9.0&2.30&20.0&2.30&9.0&3.70&0.30&0.10&7.0&9.0&87.0&0.10&249$\pm$90&29.6$\pm$10.2\\
$\overline{A}_q A_q$&1.40&4.10&1.0&7.20&1.50&3.40&19.2&4.0&8.80&6.40&0.40&0.0&10.0&2.80&65.0&27.0&220$\pm$69&87.4$\pm$29.5\\
$\overline{V}_q V_q$&5.20&0.40&1.0&0.30&6.50&0.30&11.8&1.50&5.40&2.40&1.90&0.20&9.0&0.30&0.90&0.10&102$\pm$18&17.2$\pm$2.9\\
Eq.\,\ref{eq:current}&&&&&&&&&&&&&&&&&&\\
$\overline{A}_q A_q$&3.0&3.6&1.5&2.0&4.8&2.0&37.5&7.70&17.6&12.3&0.80&0.10&12.0&7.0&108&72.0&448$\pm$117&136$\pm$74\\
\\
$M_{_{\mathcal{M},\mathcal{T}}}$[$\MeV$]&&&&&&&&&&&&&&&&&&\\
$0^{++}$ Molecule &&&&&&&&&&&&&&&&&&\\
$\overline{\chi}_{q0}\chi_{q0}$&11&39&8.0&28&10&24&47&36&19&18&29&13&76&112&9.0&8.0&6675$\pm$98&19653$\pm$131\\
$\overline{\eta}_q \eta_q$&23&4.0&3.0&15&23&26&51&29&24&49&14&13&186&58&3.8&1.6&6029$\pm$198&19259$\pm$88\\
$\overline{J\psi}J\psi,\,\overline{\Upsilon}\Upsilon$&34&31&11&42&24&27&27&52&49&30&31&22&359&116&1.3&0.0&6376$\pm$367&19430$\pm$145\\
$\overline{\chi}_{q1}\chi_{q1}$&26&4.0&29&99&20&22&42&25&20&43&5.0&22&16&73&7.0&6.0&6494$\pm$66&19770$\pm$137\\
$0^{++}$ Tetraquark &&&&&&&&&&&&&&&&&&\\
Table \ref{tab:current} &&&&&&&&&&&&&&&&&&\\
$\overline{P}_q P_q$&34&10&19&40&23&24&46&28&20&46&30&22&258&23&22&5.0&6795$\pm$268&19754$\pm$79\\
$\overline{S}_q S_q$&12&1.0&28&38&21&26&54&29&43&59&1.0&2.0&25&89&9.0&9.0&6411$\pm$83&19217$\pm$120\\
$\overline{A}_q A_q$&26&37&32&132&20&23&43&25&21&43&2.0&1.0&38&53&0.0&10&6450$\pm$75&19872$\pm$156\\
$\overline{V}_q V_q$&59&27&10&22&26&4.0&47&29&25&50&21&15&152&39&1.0&0.10&6462$\pm$175&19489$\pm$79\\
Eq.\,\ref{eq:current}&&&&&&&&&&&&&&&&&&\\
$\overline{A}_q A_q$&4.0&21&3.0&95&21&25&43&27&21&47&2.0&0.0&39&30&16&2.0&6471$\pm$67&19717$\pm$118\\
\hline
\hline
\end{tabular*}
}}
\label{tab:results}
\end{table*}
\section{Confrontation with some LO results and data}
\subsection*{$\bullet$ Comparison with some LO QSSR and MOM results}
\nin
Using the ratio of moments in Eq.\,\ref{eq:MOM} we evaluate the mass of $\overline{\chi}_{q0}\chi_{q0}$ and $\overline{S}_cS_c$:
\bea
\mathcal{M}_n(Q^{2}_{0})&=&\frac{1}{\pi}\int^{\infty}_{16m^{2}_{Q}} dt \frac{\mbox{Im}\Pi_{\mathcal{M,T}}(t)}{(t+Q^{2}_{0})^n}~,\nnb\\
M^{2}_{\mathcal{M,T}}&=&\frac{\mathcal{M}_n(Q^{2}_{0})}{\mathcal{M}_{n+1}(Q^{2}_{0})}-Q^{2}_{0}~,
\label{eq:MOM}
\eea
-- From MOM at NLO $\oplus\,\lag \alpha_s G^2\rag$, we obtain:
\bea
M_{\chi_{c0}\chi_{c0}}\simeq 6.93\, \mbox{GeV}\, ,~~M_{S_cS_c}\simeq 6.38\, \mbox{GeV},
\eea
compared to the ones from LSR in Table \ref{tab:results}, these results indicate that the predictions from the two methods (LSR and MOM) are in agreement within the error.\\
-- From MOM at LO$\,\oplus\,\lag \alpha_s G^2\rag$:
\bea
M_{\chi_{c0}\chi_{c0}}\simeq 6.78\, \mbox{GeV}\, ,~~M_{\chi_{b0}\chi_{b0}}\simeq 19.53\, \mbox{GeV},
\eea
which are lower than the ones from \cite{CCLSZ}. With the inclusion of the $\alpha_s$ QCD corrections, our LSR predictions for the charm and $\overline{P}_b P_b$ cases are in good agreement within the error with the LO ones from \cite{CCLSZ}. However, for the $\overline{S}_b S_b,\,\overline{A}_b A_b$ and $\overline{V}_b V_b$ states our results disagree. Due to the difficulty to compare the expressions of the full correlator in \cite{CCLSZ} with the spectral function we cannot trace back the discrepancy.\\
-- Using Eq.\,\ref{eq:current}, our masses predictions from LSR at LO for $\overline{A}_q A_q$:
\bea
M_{A_c A_c}\simeq 6.50\, \mbox{GeV},~~M_{A_b A_b}\simeq 19.49\, \mbox{GeV},
\eea
are lower than the one of \cite{WANG}. The estimated masses of $\overline{A}_q A_q$ from \cite{CCLSZ} are higher (resp. lower) than the ones from \cite{WANG} for the charm (resp. bottom) channel. Such discrepancies may be explained by an unusual treatment of the sum rules by the author of \cite{WANG}.
\subsection*{$\bullet$ Confrontation with experiments}
We conclude from the previous analysis that:\\
-- The broad structure around (6.2$\sim$6.7) GeV might be explained by the $\overline{\eta}_c \eta_c,\,\overline{\chi}_{c1}\chi_{c1}$ and $\overline{J/\psi} J/\psi$ molecules or/and the $\overline{S}_c S_c,\,\overline{A}_c A_c$ and $\overline{V}_c V_c$ tetraquarks.\\
-- The 	narrow structure around 6.9 GeV, if it is a $0^{++}$ state, can be identified with a $\overline{\chi}_{c0}\chi_{c0}$ molecule or $\overline{P}_c P_c$ tetraquark.\\
-- The $\overline{\chi}_{c1}\chi_{c1}$ predicted mass is below the $\chi_{c1}\chi_{c1}$ threshold, while for the beauty state all of the predicted masses are above the $\eta_b\eta_b$ and $\Upsilon(1S)\Upsilon(1S)$ thresholds.\\
-- Our predictions cannot clearly disentangle the mass of a molecule from a tetraquark state with the same quantum numbers.
\section{Conclusions}
We have presented improved predictions of QSSR for the masses and couplings of fully heavy $0^{++}$ molecules and four-quarks states at NLO  of PT series and including non-perturbative $\lag \alpha_s G^2 \rag$ and $\lag G^3 \rag$ contributions. Using our calculation method, the effect of the heavy quark condensate is included into the gluon condensate one \cite{BRG1,BLP1,BLP2}. We can see a good convergence of the PT series after including higher order corrections which confirms the veracity of our results. Our analysis has been done within stability criteria with respect to the LSR variable $\tau$, the QCD continuum threshold $t_c$ and the subtraction constant $\mu$ which have provided successful predictions in different hadronic channels \cite{BERT,BELL,NEUF,SNB1,SNB11,BERT2,BERT3,MARR,SNA2}. The optimal values of the masses and couplings have been extracted at the same value of these parameters where the stability appears as an extremum and/or inflection point. We have taken as a final result, the mean obtained with and without the $\lag G^3 \rag$ contribution and considered the error induced in this way as systematics due to the truncation of the OPE.\\
\nin In a future work, we plan to evaluate the spectra and widths of $2^{++}$ four-quark states.

\end{document}